\documentclass[review,journal,english]{vgtc}         

\ifpdf
  \pdfoutput=1\relax                   
  \usepackage{graphicx}                
  \DeclareGraphicsExtensions{.pdf,.png,.jpg,.jpeg} 
\else
  \usepackage{graphicx}                
  \DeclareGraphicsExtensions{.eps}     
\fi%

\graphicspath{{figures/}{pictures/}{images/}{./}} 

\usepackage{microtype}                 
\PassOptionsToPackage{warn}{textcomp}  
\usepackage{textcomp}                  
\usepackage{mathptmx}                  
\usepackage{times}                     
\usepackage{cite}                      
\usepackage{tabu}                      
\usepackage{booktabs}                  
\usepackage{soul}
\usepackage[markup=underlined]{changes}
\usepackage[color=red!40]{todonotes}

\usepackage{pifont}
\usepackage{epstopdf}
\usepackage{times}
\usepackage{amssymb}
\usepackage{enumitem}
\usepackage{multirow}
\usepackage{booktabs}
\usepackage{colortbl}
\usepackage[super]{nth}
\usepackage{url}
\usepackage{lipsum, babel}

\newcommand{\name}{ICE} 
\newcommand{\etal}{et~al.\ }
\newcommand{\eg}{e.g.}
\newcommand{\ie}{i.e.}

\newcommand{\siwei}[1]{\textcolor{black}{#1}}

\newcommand*{\rom}[1]{\uppercase\expandafter{\romannumeral #1\relax}}

\expandafter\def\expandafter\normalsize\expandafter{%
    \normalsize
    \setlength\abovedisplayskip{8pt}
    \setlength\belowdisplayskip{8pt}
    \setlength\abovedisplayshortskip{8pt}
    \setlength\belowdisplayshortskip{8pt}
}

\newlength\colEventName 
\newlength\colEventDesc 
\newlength\colEvent 
\newlength\colEventNote 

\onlineid{1066}

\vgtccategory{Application}
\vgtcpapertype{Application/Design Study}

\title{\name: \textul{I}dentify and \textul{C}ompare \textul{E}vent Sequence Sets \\through Multi-Scale Matrix and Unit Visualizations}

\author{Siwei Fu, Jian Zhao, Zhicheng Liu, Kwan-liu Ma, and Huamin Qu}
\authorfooter{
\item
 S. Fu and H. Qu are with Hong Kong University of Science and Technology. E-mail: \{sfuaa, huamin\}@cse.ust.hk.
\item
 J. Zhao is with FX Palo Alto Laboratory. E-mail: zhao@fxpal.com. 
\item
 Z. Liu is with Adobe Research. E-mail: leoli@adobe.com.
\item
 K. Ma is with UC Davis. E-mail: ma@cs.ucdavis.edu.
}

\shortauthortitle{Biv \MakeLowercase{\textit{et al.}}: Global Illumination for Fun and Profit}

\abstract{Comparative analysis of event sequence data is essential in many application domains, such as website design and medical care. 
However, analysts often face two challenges: they may not always know which sets of event sequences in the data are useful to compare, and the comparison needs to be achieved at different granularity, due to the volume and complexity of the data. 
This paper presents, \name{}, an interactive visualization that allows analysts to explore an event sequence dataset, and identify promising sets of event sequences to compare at both the pattern and sequence levels. 
More specifically, \name{} incorporates a multi-level matrix-based visualization for browsing the entire dataset based on the prefixes and suffixes of sequences. To support comparison at multiple levels, \name{} employs the unit visualization technique, and we further explore the design space of unit visualizations for event sequence comparison tasks. 
Finally, we demonstrate the effectiveness of \name{} with three real-world datasets from different domains.

} 

\keywords{Event sequence data, visual comparison, matrix-based visualization, unit visualization.}

\CCScatlist{ 
 \CCScat{K.6.1}{Management of Computing and Information Systems}%
{Project and People Management}{Life Cycle};
 \CCScat{K.7.m}{The Computing Profession}{Miscellaneous}{Ethics}
}

\teaser{
  \centering
  \includegraphics[width=.9\linewidth]{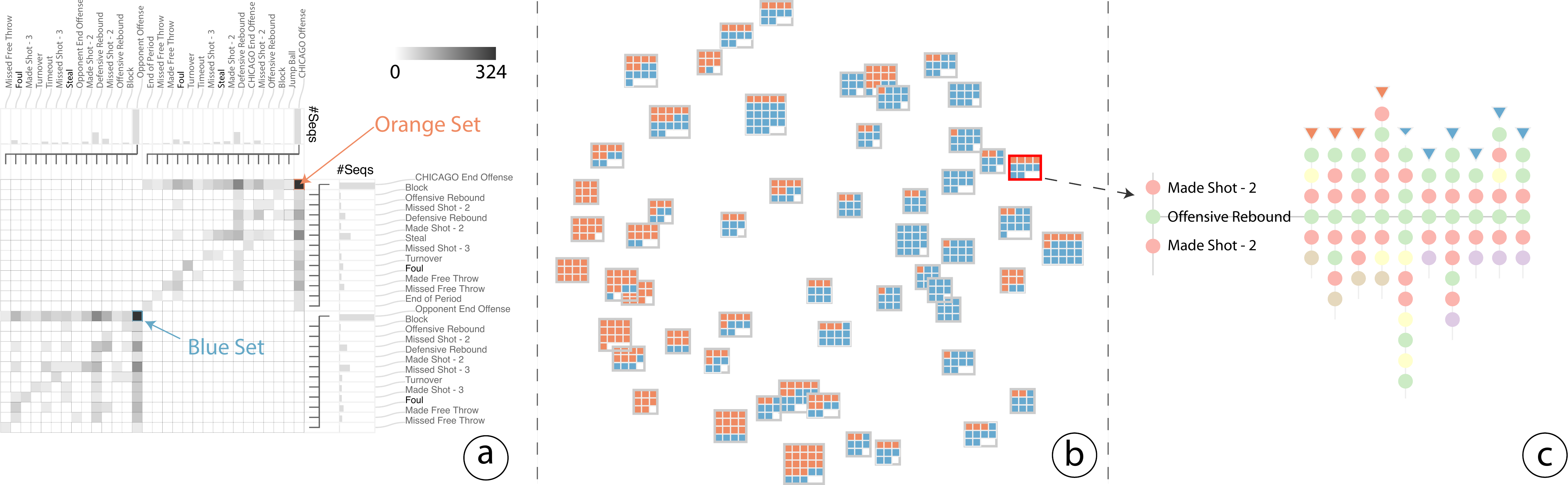}
  \caption{
  The workflow of using \name{}. (a) After loading a dataset, an analyst can explore all the event sequences according to their prefixes and suffixes using a multi-scale matrix-based visualization. 
  (b) She can then identify two interesting sets of sequences by selecting the cells in the matrix. Sequential patterns are calculated on the fly and displayed based on unit visualization techniques. 
  (c) Further, an analyst can dig into a pattern of interest to investigate details of the sequences that manifest the pattern.
  }
 \label{fig:teaser}
}

\begin{document}


\firstsection{Introduction}

\maketitle




Event sequence data is commonly found in many areas.  For example, in health care, electronic health records (EHR) store the symptoms, treatments, and outcomes of each patient in a temporal order~\cite{Wongsuphasawat2012}; and in website management, users' browsing behaviors at each page are logged as sequences, \ie, clickstreams~\cite{Liu2017a}. 
In practice, analysts often seek to compare different sets of event sequences to obtain insights. 
Consider two event sequences of patients' symptoms with different outcomes: death and recover; doctors would like to know what are the key symptoms leading to the results?
Also, consider users' clickstreams on an e-commerce website; web analysts may want to find out what are the key characteristics of behaviors between users who started with different landing pages?

It is often challenging to perform such comparison tasks mainly due to two issues. 
The first one is related to ``What to compare?'' In many cases, an analyst may not know exactly which two sets of sequences are meaningful to compare in a dataset. It is an iterative and exploratory process for locating interesting subsets of data for comparison, especially when an analyst is unfamiliar with the data. Nonetheless, most related systems assume users understand the data well and have clear purposes about what to compare~\cite{Hibino1997,Monroe2013a,Jin2010,Krause2016}.

The second challenge is related to ``How to compare?'' As tasks and goals are diverse in practice, the comparison is usually needed at different granularity with flexibility. At a higher level, an analyst may be interested in how frequent patterns differ in two sets of sequences; and at a lower level, raw sequences are explored and compared in detail. But existing systems mostly focus on comparing either raw sequences~\cite{Wongsuphasawat2009,Du2016,Zhao2015,Malik2015} or patterns~\cite{Polack2015}, lacking the support of flexible comparison from different levels and perspectives.

To address these challenges, we present \name{}, an interactive visualization that allows analysts to explore an event sequence dataset and identify sets of sequences to compare at different granularity including frequent patterns and raw sequences.
Inspired by Vehlow et~al.'s work~\cite{Vehlow2016}, we propose a multi-scale matrix-based visualization with its rows and columns corresponding to the nodes of the prefix and postfix trees constructed from the dataset, and thus each cell in the matrix represents sequences starting and ending with specific events in particular orders. This provides a visual summarization of the entire dataset, so an analyst could explore sequences by tracing them from both directions and select interesting subsets to compare.
We employ the unit visualization technique~\cite{Park2017} to support the comparison of two event sequence sets at different levels, and explore the design space of applying such technique in comparison tasks. This approach offers the flexibility of grouping and positioning the basic units (\ie, each unit is a sequence) in many ways, such as grouped by common patterns and placed in 2D based on similarity, to serve an analyst's diverse comparison goals.

To evaluate \name{}, we conduct three case studies with real-world datasets in different domains, including action event sequences in football matches, students' interaction logs of using a course software system, and user clickstreams on an E-commerce website.
The results indicate that \name{} helps analysts effectively explore an event sequence dataset, identify interesting sets of sequences, and compare two sets at multiple granularities.

\section{Related work}
Our work is closely related to research areas in visual summarization, query, and comparison of event sequences. Besides, our design is inspired by the unit visualization. We review the state of the art of these aspects in order.

\subsection{Visual summarization of event sequence datasets}
To obtain a high-level overview of common paths and their volumes in an event sequence dataset, some researchers extracted tree structures and illustrated them using icicle plots or node-link diagrams~\cite{Wongsuphasawat2011,Liu2017,Shen2012,Monroe2013,Kruskal1983}. 
Others attempted to consolidate the whole dataset into a transition graph to demonstrate the flows of events, for example, OutFlow~\cite{Wongsuphasawat2012}, CareFlow~\cite{Perer2013}, and DecisionFlow~\cite{Gotz2014}.

An increasing number of visualization systems incorporates analytical approaches to understand common patterns shared by sequences.
FP-Viz is one of the early systems, which visualizes frequent patterns using a Sunburst diagram~\cite{Leung2009}. 
Wang \etal\cite{Wang2016} and Vrotsou \etal\cite{Vrotsou2009} mined dominating user behaviors from clickstreams data and further presented them using intuitive visualization, such as node-link diagram and circle packing layout.
In addition, Wei \etal proposed an overview of clusters in clickstreams data with a Self-Organizing Map~\cite{Wei2012}.
Similarly, WebCANVAS~\cite{Cadez2003} and LogView~\cite{Makanju2008} are designed to show the hierarchical structures generated by sequence clustering methods using tree visualization techniques, such as TreeMaps\cite{Shneiderman1992} and node-link diagrams.
Recently, Liu \etal proposed a three-stage analysis pipeline for event sequence analysis~\cite{Liu2017a}. 
In their work, patterns are displayed using glyphs and sequences can be aligned with a particular event.

The design of \name\ has been inspired by many of the above visualizations. But we focus on the comparative analysis of event sequence data. 
Although comparing two sets of sequences could be supported by creating two instances of these techniques, the process may be cumbersome as they are not designed specifically for comparison tasks.

\subsection{Visual query of event sequences}
The increasing volume of event sequence data has created an overarching need for querying meaningful records. 
Several techniques are proposed to allow users to define queries based on interval, event absence, and other temporal constraints of events~\cite{Hibino1997,Monroe2013a,Jin2010,Krause2016}. 
However, these approaches do not support querying by starting or ending events. 
To address this, Outflow~\cite{Wongsuphasawat2012} and CareFlow~\cite{Perer2013} offer the specification of ending events and the exploration of pathways associated with these results. 
DecisionFlow moves one step forward to support querying with preconditions (starting events) and outcomes (ending events)~\cite{Gotz2014}.
To augment expressiveness, Eventpad~\cite{Cappers2018} and (s$\vert$qu)eries~\cite{Zgraggen2015} enable users to visually construct regular expressions for sequences query.

These techniques are effective when analysts know what events to specify in their queries. However, in many scenarios, they lack such knowledge about the data, and visual exploration is needed to identify meaningful sets of sequences to compare.
\name{} addresses this issue by offering a matrix-based visualization to provide a visual summarization of the entire data from the perspectives of starting and ending events.

\subsection{Visual comparison of event sequences}

Although the concept of visual comparison is not new (see \cite{Gleicher2011} for visual comparison techniques), applying this to event sequences is generally under-exploited in the literature.
There exists some work targeting at comparison of individual sequences. For example,
Similan~\cite{Wongsuphasawat2009} and EventAction~\cite{Du2016} employ certain metrics to identify similar sequences and present the sequences side by side. 
Another example is TimeSlice~\cite{Zhao2012}, which provides exploration and comparison of faceted event sequences. Although useful, these techniques are not adequate for comparing two sets of sequences. Because they focus on event details in individual sequences, and cannot be generalized to compare sequence sets. 
To support such comparison, MatrixWave~\cite{Zhao2015} utilizes a series of transition matrices, and CoCo~\cite{Malik2015,Malik2016} compares sequence sets based on some predefined metrics, such as the number of records, number of events, and prevalence of an event. 
Further, TimeStitch defines cohorts by frequent patterns and compares two cohorts side by side~\cite{Polack2015}.

The above techniques focus on the comparison of either raw sequences or frequent patterns, whereas \name{} supports comparing sets of sequences at different granularity. With much flexibility, individual sequences can be organized in patterns and displayed in a range of layouts, and can be then further explored by analysts.

\subsection{Unit visualizations}

According to Part \etal\cite{Park2017}, unit visualization is defined as ``visualization that maintains the identity property of its visual marks, \ie, where each visual mark is a unique entity that is associated with a corresponding unique data item.'' The unit visualization has many benefits, including semantic constancy, direct interaction, smooth animation, etc~\cite{Park2017}.
Many systems based on this approach have been proposed in recent years, such as SandDance~\cite{Drucker2015}, Squares~\cite{Ren2017}, Gatherplots~\cite{Park2016}, and Past Visions~\cite{Glinka2016}.
To enhance the unit visualization, Oelke \etal discussed different approaches for visual boosting~\cite{Oelke2011}.
To formalize the creation of unit visualizations, Drucker and Fernandez characterized the design space of unit visualization and proposed a unifying framework~\cite{Drucker2015}.
Similarly, Park \etal proposed a grammar for unit visualizations named ATOM~\cite{Park2017}. The expressive power of ATOM enables it to describe unit visualizations with various complexity. 

In this paper, we have explored the design space of unit visualizations for comparing two sets of sequences.
Our design space borrows the spatial layouts of unit visualizations in ATOM~\cite{Park2017}, and moves one step further by adapting the layouts to visual comparison tasks based on Gleicher's schema~\cite{Gleicher2011}.

\section{Task analysis} 
\label{sec:task}

We aim to design a visual analysis tool to facilitate the comparison of different subsets of sequences discovered based on the visual exploration of the entire event sequence dataset. We follow Brehmer and Munzner's typology~\cite{Brehmer2013} to derive proper analytical tasks by expressing them as a series of basic tasks, and articulating the \textit{why} and \textit{how}.

\begin{description}

\item \textbf{T1: Explore event sequences based on prefixes and/or suffixes.}
\begin{itemize} 
\itemsep 0mm

\item \textit{Why: discover $\rightarrow$ locate + browse + explore $\rightarrow$ identify}

Analysts often seek to understand the whole dataset from the perspectives of starting (prefix) and ending (suffix) events. For example in web analytics, a web analyst may ask questions like:
``What are the most popular entry points to our website in this marketing campaign,'' and 
``For customers brought by the search engine \textit{X}, how many of them check out finally?''~\cite{Zhao2015}
However, in many situations, analysts lack the preliminary knowledge about the data and may not know what to compare. Thus, they need to \textit{discover} insights with the help of visualization, in which they have to \textit{locate}, \textit{browse}, and \textit{explore} the sequences because the location and target are unknown. The general goal is to \textit{identify} interesting sets of sequences, which is tied to \textbf{T2}.

\item \textit{How: aggregate + encode + arrange}

To support this task, the visualization should \textit{aggregate} all sequences based on their prefixes and suffixes to enable effective exploration within the whole dataset.
Moreover, the visualization should \textit{encode} and \textit{arrange} the aggregated data in a meaningful way (\eg, sorted by some criteria). 
\end{itemize}

\item \textbf{T2: Identify interesting sets of event sequences for comparison.}
\begin{itemize} 
\itemsep 0mm

\item \textit{Why: discover $\rightarrow$ look up + browse $\rightarrow$ compare}

Based on the starting and ending events, analysts could \textit{discover} interesting sets of sequences in three ways: by the prefix, the suffix, and both~\cite{Gotz2014}. This process is multi-scale in nature, in which analysts would like to \textit{browse} sequence sets with different prefix and/or suffix lengths, and \textit{look up} the events from the start and/or the end.
For example, in clickstreams on E-commerce websites, analysts may be interested in the sequences ending with ``Checkout'' followed by ``Error'' to diagnose the problems that users had after checkout. 
Overall, the goal is to find candidate sets of sequences to \textit{compare}, which leads to \textbf{T3} and \textbf{T4}.

\item \textit{How: navigate + filter + select}

To enable this action, the visualization should help analysts \textit{navigate} the whole dataset at multiple levels, both from the starts and ends of the sequences. Further, due to the overwhelming data volume and the large space of event type permutations, the visualization must allow users to \textit{filter} out parts that are not considered for comparison. This filtering could be supported based on specific events that analysts \textit{select} during their multi-scale visual exploration.

\end{itemize}

\item \textbf{T3: Compare two sets of event sequences at pattern level.}
\begin{itemize}
\itemsep 0mm

\item \textit{Why: discover $\rightarrow$ search $\rightarrow$ summarize + compare}

After analysts identify two sequence sets of interest, one important task is to \textit{discover} high-level insights by comparing their frequent patterns mined by algorithms, such as VMSP~\cite{Fournier-Viger2014a} and SPAM~\cite{Ayres2002}.
Example questions that analysts may raise include: ``How do patterns differ between two sets,'' ``Which pattern clusters are dominated by one set or the other,'' and ``Do these patterns form some clusters?''
To answer these questions, analysts need to conduct visual \textit{search} of the patterns from two different sets of sequences in order to \textit{summarize} and \textit{compare} them.

\item \textit{How: arrange + aggregate + encode + change}

The visualization should therefore allow analysts to view an ensemble of the diverse patterns, which should be \textit{arranged} and \textit{aggregated} to reflect their relationships and similarity. Techniques such as clustering, and sorting can be applied. 
To reflect the nuances among the patterns and the sequences containing them (that may be from two different sets), some quantities, such as the support of the mined patterns and the proportion of sequences from each set, should be \textit{encoded} to ease the comparison at the pattern level. Also, the visualization should offer an ability to \textit{change} the representation, such as arrangement and encoding methods, for different comparison goals.

\end{itemize}

\item \textbf{T4: Compare raw sequences exhibiting particular patterns.}
\begin{itemize}
\itemsep 0mm

\item \textit{Why: discover $\rightarrow$ browse + locate $\rightarrow$ compare}

In addition to comparing two sets of sequences at a higher level, analysts may want to dig into the raw sequences that contain a specific pattern of interest~\cite{Liu2017a}.
Thus, they would like to \textit{discover} the connections between the patterns and sequences, where they need to \textit{browse} events in the sequences and \textit{locate} individual events with significance, with the ultimate goal to \textit{compare} the sequences, for example, based on key events.

\item \textit{How: select + arrange + change}

To achieve this task, the visualization should allow analysts to \textit{select} a specific pattern and \textit{arrange} the associated sequences according to events in the pattern or other key events of interest (\eg, aligning all sequences based on a particular event).
To assist the comparison, analysts should be able to \textit{change} the events of interest to alter the alignment of the sequences. 

\end{itemize}

\end{description}



\section{Design space of sequence set comparison}
\label{sec:designspace}
Assuming that two sets of sequences are identified in the data, a major question is how to compare them effectively at different granularity, as mentioned by \textbf{T3} and \textbf{T4}. Such comparison is the ultimate goal of an analysts using the visualization tool in our scenario.
Since a large body of work (\eg, \cite{Wongsuphasawat2009,Du2016,Zhao2015}) has focused on comparison at the sequence level, in this section, we discuss the design space of visualization techniques that address the pattern level comparison of sequences.
Following the notations by Liu~\etal\cite{Liu2017a}, we first describe our data model and then introduce the design space.

\subsection{Data model}

An \textit{event sequence dataset} $D$ contains a number of sequences $S_i$. That is, $D=\{S_1, S_2, \cdots, S_s\}$. 
A \textit{sequence} $S$ is defined as a set of ordered events: $S=\{E_1, E_2, \cdots, E_n\}$. 
Each \textit{event} $E$ could be multivariate. For example, a event has a ``name'' to present its identity and a ``timestamp'' to record when it happened, etc.
A \textit{set of patterns} $\mathcal{P}=\{P_1, P_2, \cdots, P_s\}$ can be obtained by applying the sequential pattern mining algorithm to the dataset $D$. 
Each \textit{pattern} $P$ is a series of events contained in one or more sequences. That is, $\exists \{S_1, S_2, \cdots, S_k\}\subseteq D,\, k>0$, such that $P\sqsubseteq S_1, P\sqsubseteq S_2, \cdots, P\sqsubseteq S_k$.
Each pattern $P$ is associated with a \textit{support set}, \ie, $supset(P)$, which is a set of sequences in $D$ that contain the pattern: $supset(P)=\{S|S\in D, P\sqsubseteq S\}$.
The \textit{support} of a pattern $supp(P)$ is the ratio of $|supset(P)|$ and $|D|$: $supp(P)=\frac{|supset(P)|}{|D|}\times 100\%$.

In our implementation, we employ the VMSP algorithm~\cite{Fournier-Viger2014a} for mining maximal sequential patterns; however, \name{} is designed to work with any sequential pattern mining algorithms for extracting any kinds of patterns, such as frequent patterns~\cite{Ayres2002} and closed patterns~\cite{Fournier-Viger2014a}.
After an analyst identifies two sets of sequences to compare, $D_1$ and $D_2$, we join the two sets and apply the VMSP algorithm to obtain the patterns $\mathcal{P}=\{P_1, P_2, \cdots, P_s\}$. The support set of each pattern $supset(P)=\{S_1, S_2, \cdots, S_k\}$ contains sequences from either $D_1$ or $D_2$.


\subsection{Design space of pattern level comparison}

When comparing two sets of sequences at the pattern level, it is essential to reveal the relationships between the mined patterns and the sequences containing a particular pattern (\ie, the support sets). This facilitates the ``Overview first, zoom and filter, and details on demand'' approach~\cite{} of conducting comparison tasks.
To allow an analyst to obtain an effective overview of the patterns while accessing some information about the support sets, we employ the unit visualization technique~\cite{Park2017} where each sequence is mapped to a visual mark (\ie, the basic unit) and a pattern is a group of corresponding visual marks. 


\subsubsection{Expressing design space with ATOM}

\begin{figure}[tb]
 \centering
 \includegraphics[width=\linewidth]{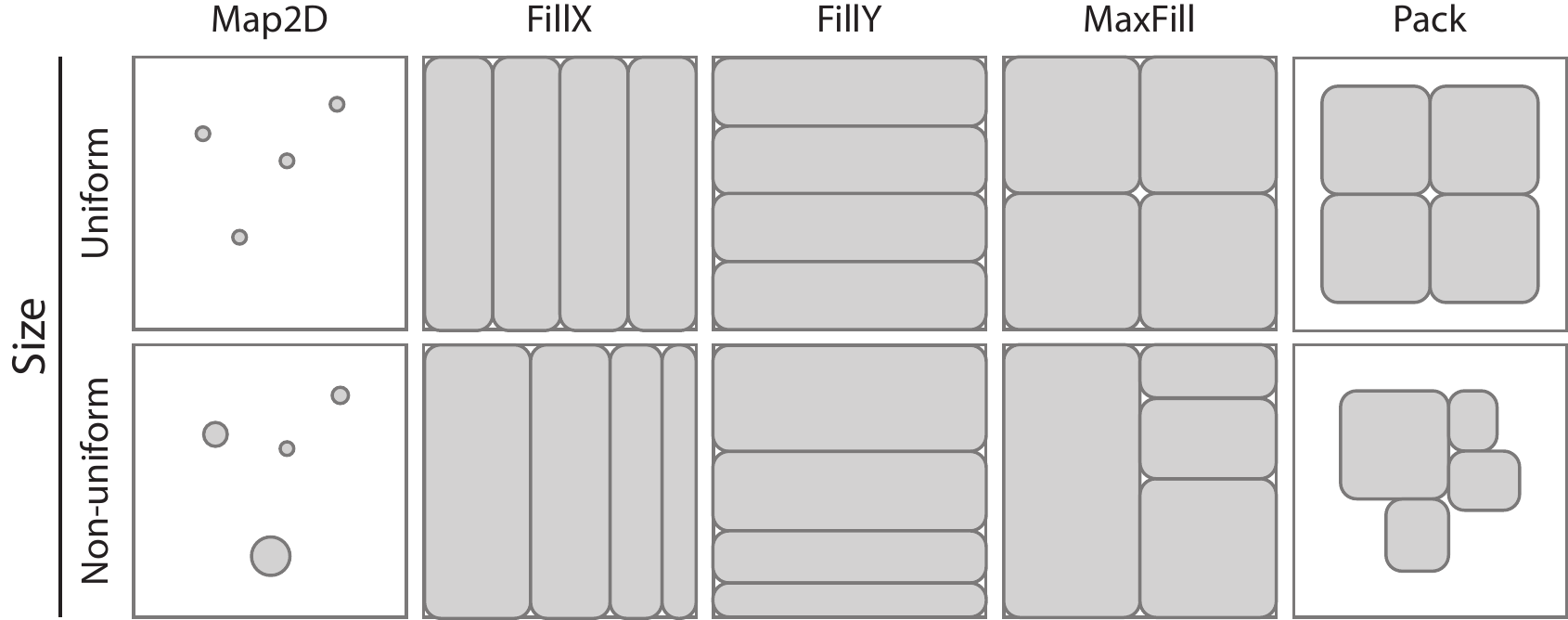}
 \caption{The visual representation of five layout operations in the spatial domain in ATOM~\cite{Park2017}. The size of sub-containers (in gray) can be uniform or non-uniform (\ie, mapped to a value such as the count of data items).}
 \label{fig:layout}
\end{figure}

ATOM is a visual grammar for systematically describing the spatial arrangement of each visual mark (that represents a data item) in unit visualizations~\cite{Park2017}.
The grammar defines a container as the composition of a \textit{dataset} and a \textit{canvas}.
Hence, a root container consists of the entire dataset and all the visual space available. 
Then, we recursively apply the unit visualization \textit{layout operations} until all containers areF associated with only one data item.
Especially, the operations manipulate containers in both the data domain and the spatial domain. To be specific, in the data domain, the operations 
``divides a dataset of parent container into a set of datasets for child containers,''
and in the spatial domain, the operations ``split the parent space into child spaces.''~\cite{Park2017}
Different layout operations in the spatial domain are demonstrated in Figure~\ref{fig:layout}.

In our case, the root container consists of the entire dataset, \ie, all the patterns detected from the sequential pattern mining, and the whole canvas. 
We notice that the data domain is hierarchical in nature. Thus, we assign a sub-container for each pattern in the dataset; therefore, each sub-container consists of the corresponding support set and the visual space allocated by the upper level.
We can apply any of the layout operations, including \textit{Map2D}, \textit{FillX}, \textit{FillY}, \textit{MaxFill}, and \textit{Pack}, to both the root container and sub-containers, which form the design space of visualizing the patterns and their related sequences (Figure~\ref{fig:designspace}). As FillX and FillY are similar, we group them in our design space for simplicity.

Particularly, we employ non-uniform layouts at the pattern level (\ie, within the root container), where the area of the visual space of a pattern is proportional to the number of sequences in its support set.
This allows for more screen real estate for more frequent patterns in order to display the sequences associated with the pattern. If not, each sequence may be too small due to the limitation of unit visualizations that every unit needs to be displayed and distinguished~\cite{Park2017}.
At the sequence level (\ie, within a sub-container), we use uniform layouts with shared size of visual marks. This is because the geometrical properties of the visual representations of all sequences are the same, which facilitates the comparison across different patterns.
However, our discussion of the design space (see Section~\ref{subsubsec:designalter}) could be generalized to other configurations, such as uniform layouts at both levels.


\subsubsection{Cooperating with visual comparison}
Based on the above design space described with ATOM, we consider the basic visual comparison approaches summarized by Gleicher \etal\cite{Gleicher2011}, including juxtaposition, superposition, and explicit encoding, in order to accommodate the comparison of two sequence sets with unit visualizations.

\begin{figure}[tb]
 \centering
 \includegraphics[width=\linewidth]{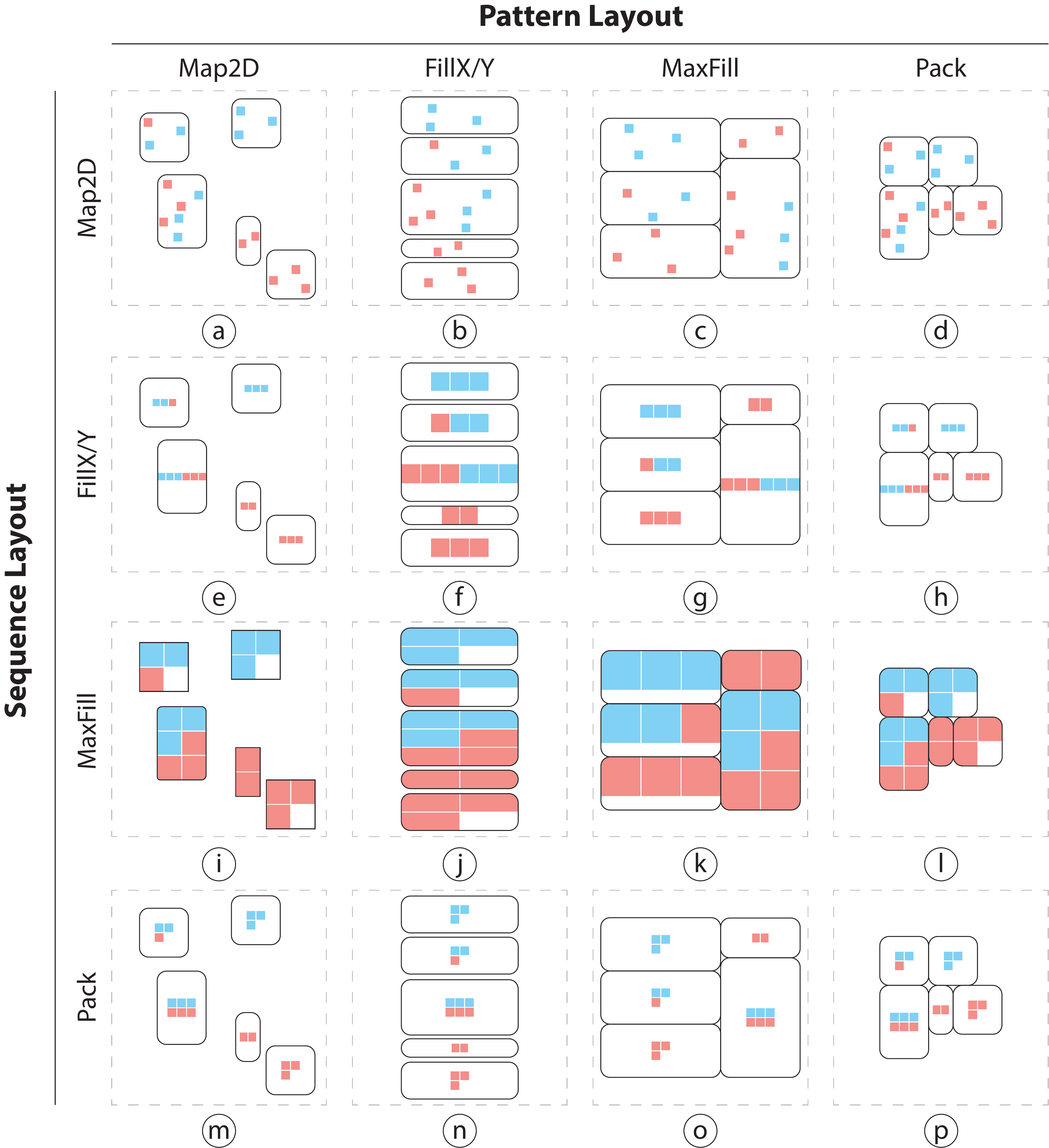}
 \caption{The design space of employing unit visualizations for the comparison of two event sequence sets at the pattern level. The design alternatives are demonstrated with a hypothetical dataset where four unique patterns are detected from the two sets of sequences for comparison. 
 In each design alternative, the dashed border rectangle represents the entire visual space. Each rounded rectangle represents the specific visual space allocated to that pattern. Sequences exhibiting that pattern are shown as visual marks color coded by their memberships to the two sets.
 } \label{fig:designspace}
\end{figure}

As shown in Figure~\ref{fig:designspace}, we finally employ superposition, so patterns detected from both sets are merged and displayed on the same canvas, and sequences within each pattern (\ie, the support set) are color-coded based on which set they belong to. If a sequence appears in multiple patterns, we simply duplicate that sequence in our visualization.
 
If juxtaposition is applied, sequences from two different sets are isolated and located on two visualizations side by side, which may hinder an analyst from interpreting and comparing the data because they need to visually search and match the patterns of two sets on two canvases.
For explicit encoding, a difference quantity needs to be computed and visually represented, in our case, the difference between two patterns, or two support sets. However, it is difficult to define the difference as a numerical variable to encode visually, because each support set contains multiple sequences that could be dramatically different. Although some simple variables can be calculated such as the size difference of support sets, it is not informative for comparison. 
However, future empirical studies need to be conducted to verify the above intuitions.

\begin{figure*}[!htb]
 \centering
 \includegraphics[width=\linewidth]{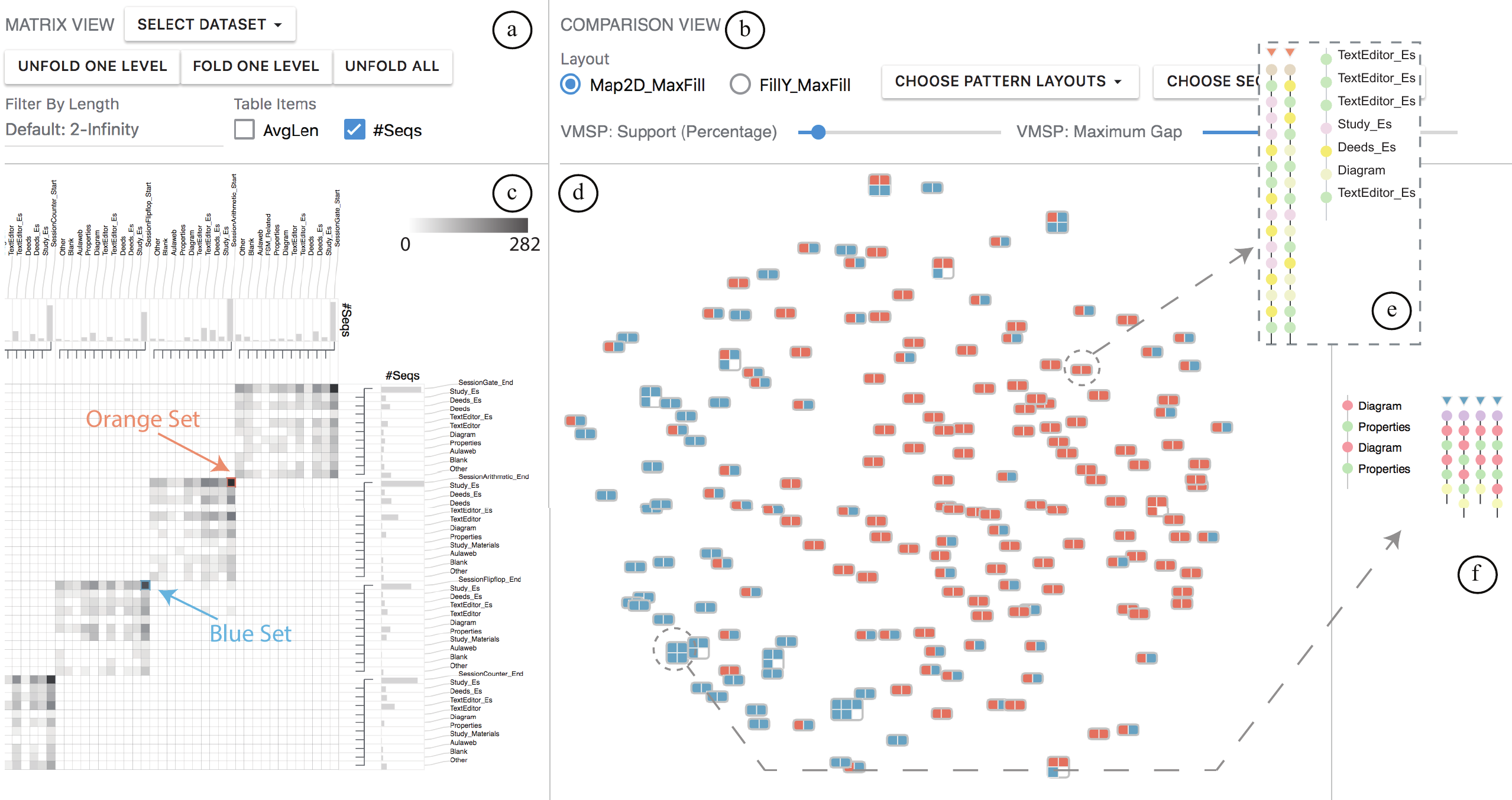}
 \caption{
 The interface of \name{} consists of two main views. A Matrix View demonstrates the entire event sequence dataset from the perspectives of prefixes and suffixes (a)(c). A Comparison View illustrates the differences between two sets of sequences at different granularity, including the pattern level (b)(d) and the sequence level (e)(f).
 } \label{fig:interface}
\end{figure*}

\subsubsection{Discussion on design alternatives}
\label{subsubsec:designalter}
Followed by Figure~\ref{fig:designspace}, here we discuss how each design alternative in the design space can facilitate the visual comparison of two sequence sets at the pattern level. 

For the Map2D layout, patterns or sequences can be positioned based on a similarity metric (\eg, Levenshtein distance~\cite{Yujian2007}) using the MDS projection~\cite{Kruskal1978}, or based on two attribute axes like a scatterplot. This is beneficial to reveal the relationships among all detected patterns (see the 1st column of Figure~\ref{fig:designspace}.
When applying Map2D to the sequence level, a shared coordinate system is needed across all the pattern sub-containers in order to facilitate the comparison of different patterns (see the 1st row of Figure~\ref{fig:designspace}). However, the MDS algorithm is not applicable because each 2D projection of sequences in a support set could be different. 
The scatterplot method requires two meaningful attributes of the sequences as the axes, and a shared scale is necessary among all pattern sub-containers. However, some event sequence data may not contain such metadata. Further, it worth noting that such sequence level Map2D layout could be more effective when the uniform layout is employed at the pattern level (Figure~\ref{fig:layout}).

In FillX or FillY, patterns or sequences are aligned in one direction which enables ordering and sorting based on certain criteria (\eg, the size of support set at the pattern level). Note that we employ non-uniform layout for patterns (see the 2nd column of Figure~\ref{fig:designspace}) and uniform layout with shared size for sequences (see the 2nd row of Figure~\ref{fig:designspace}), so some pattern sub-containers may have unfilled space~\cite{Park2017}.
There exist four different ways by applying FillX or FillY to either the pattern or the sequence level. Figure~\ref{fig:designspace} only demonstrates the configuration of FillY for patterns and FillX for sequences.
However, in any configuration, filling along one direction is not scalable for a large number of data items, where the visual representation of one item could be too small. Further, at the sequence level when the size is shared, much visual space would be wasted.

The MaxFill layout tries to utilize all the visual space, and there exist several space-filling methods such as the TreeMap~\cite{Wongsuphasawat2009} and grid-based methods. We employ the TreeMap for MaxFill at the pattern level because of its non-uniform space allocation (see the 3rd column of Figure~\ref{fig:designspace}). Although more visually compact, the TreeMap is less informative because the inter-pattern relationship or the ordering feature is missing compared to Map2D or FillX/Y.
Further, we use the grid-based approach at the sequence level to facilitate the visual comparison, since the sequences could have a common alignment (see the 3rd row of Figure~\ref{fig:designspace}). Similar to FillX/Y, due to the shared size configuration, some pattern sub-container may not be fully filled, but MaxFill is more efficient in terms of best utilizing the visual space. It can also reflect a certain order of the sequences determined by the positioning algorithm in MaxFill.

The last layout is Pack that also exhibits a family of methods such as packing compactly in 2D~\cite{Wang2006}, in 1D~\cite{}, or by a grid system~\cite{Ren2017}. As the size of pattern sub-containers is non-uniform, we apply the method by Wang \etal\cite{Wang2006} for Pack at the pattern level (see the 4th column of Figure~\ref{fig:designspace}), and use the grid-based packing at the sequence level, again, to facilitate the cross-pattern comparison of sequences (see the 4th row of Figure~\ref{fig:designspace}). 
Similar to MaxFill at the pattern level, Pack is also less informative compared to Map2D or FillX/Y. Moreover, at the sequence level, Pack may be less space efficient than MaxFill because it does not always maximize the size of visual marks.


In summary, each design alternative has its own advantages and disadvantages, which should be carefully considered case by case according to the characteristics of tasks and data. In general, based on the above discussion, the design alternatives shown in Figure~\ref{fig:designspace}(i) and (j) may be the most effective in our scenarios. 
Further, when the number of detected patterns is large, Figure~\ref{fig:designspace}(j) could be less effective.
If a shared coordinate system can be built for positioning sequences of different patterns (\ie, support sets), Figure~\ref{fig:designspace}(a) and (b) could be good alternatives, but may subject to the constraint of non-uniform pattern level layouts.
However, empirical studies for comparing user performance with different design alternatives are needed to further confirm this conclusion.

\section{\name{} interface}
\label{sec:design}




In this section, we describe the interface design of \name{} in details (Figure~\ref{fig:interface}). We first introduce the Matrix View that offers visual summarization and exploration of an event sequence dataset, and then the Comparison View that is used to compare two sets of sequences at multiple granularities.

\subsection{Matrix View: getting the gist}
As shown in Figure~\ref{fig:interface}(a)(c), the Matrix View aims to provide a summarization for the entire dataset based on prefixes and/or suffixes (\textbf{T1}), and help an analyst identify two meaningful sets of sequences for comparison (\textbf{T2}).

\begin{figure*}[tb]
 \centering
 \includegraphics[width=\linewidth]{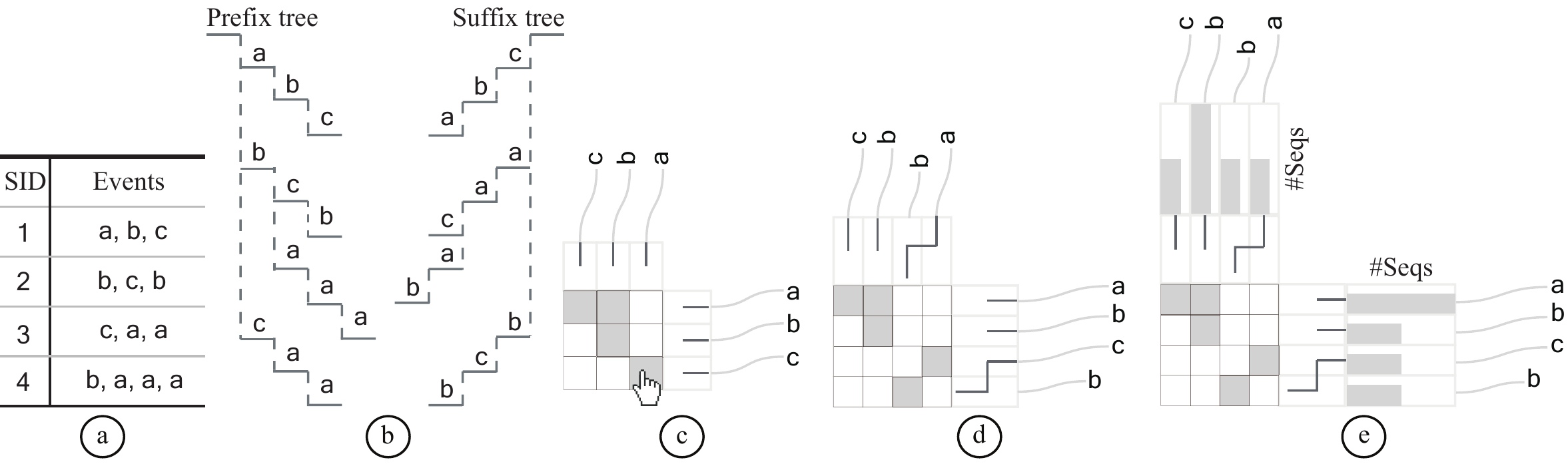}
 \caption{Illustration of the Matrix View in \name. (a) An example dataset with four sequences. (b) A prefix tree and a suffix tree are built based on the data. (c)(d) The multi-scale matrix is constructed by combining the prefix tree nodes as columns and the suffix tree nodes as rows. (e) Two bar charts indicating statistical metrics of each column and row (\eg, the total number of sequences in a row/column) are placed adjacently to the matrix.}
 \label{fig:tree}
\end{figure*}

\subsubsection{Visual encoding}
By using tree-like representations, some systems support the visual exploration of event sequences according to prefixes~\cite{Wongsuphasawat2011,Liu2017,Shen2012,Monroe2013}, so an analyst can track sequences by a series of events from the beginning.
In contrast, \name{} allows for the exploration from both the start and the end of sequences (\textbf{T1}). To do so, we first construct a prefix tree and a suffix tree based on the entire dataset. Figure~\ref{fig:tree}(a) and (b) illustrate a simple example of this process. 
Next, we construct a matrix by combining the two trees, where the columns correspond to the prefix tree nodes and the rows correspond to the suffix tree nodes. Thus, each cell in the matrix denotes all the sequences with the same starting and ending events determined by the trees (Figure~\ref{fig:tree}(c)(d)(e)). 
Some statistical quantities of sequences in each cell can be visually encoded, which allows an analyst to browse the distribution of all sequences based on their prefixes and suffixes. In our implementation, we map the number of sequences of a cell to its color density.

When the cardinality of events is large and sequences are lengthy, the size of the matrix grows exponentially. To avoid information overload, \name{} supports dynamic expanding and collapsing the trees.
For example, Figure~\ref{fig:tree}(d) shows the result after expanding the cell (a, c) in Figure~\ref{fig:tree}(c). The prefixes and suffixes are displayed with indentation to show the hierarchical structure.
Further, to help an analyst understand the characteristics of each column and row, some statistical values can be visualized at the periphery space of the matrix. In \name{}, we implemented two metrics including the number of sequences and the average sequence length~\cite{Malik2015}, where an analyst can choose to show one metric at a time as bar charts (Figure~\ref{fig:tree}(e)). Other metrics can be easily added based on real demand.

\subsubsection{User interaction}

The Matrix View provides various interaction techniques for exploring and identifying meaningful sequence sets at multiple levels based on the sequence prefixes, suffixes, and both (\textbf{T2}).

\textit{Expanding and collapsing.}
Besides clicking one cell to expand the prefix tree and suffix tree simultaneously, an analyst can click the bars outside the matrix to expand just one column or row (Figure~\ref{fig:tree}(e)). 
To facilitate multi-scale analysis, three quick access buttons are placed at the top of the Matrix View to allows an analyst to unfold and fold all nodes at the next level for both trees (Figure~\ref{fig:interface}(a)).
Unfolding or folding all nodes of the two trees are also available.

\textit{Filtering and sorting.}
An analyst could filter sequences according to their length. This is useful for removing some abnormal sequences that are extremely long or short. 
By default, \name{} shows sequences with length greater than two, which can be further adjusted using a text box.
The columns and rows of the matrix can be sorted based on a metric, for example, the total number or the average length of all sequences in a row or column.
Since the rows and columns correspond to two trees, the sorting is performed locally to keep the hierarchical structure. That is, only sibling tree nodes are sorted among each other.


\textit{Linking, zooming \& panning, and selecting.}
When an analyst hovers over a matrix cell, 
the corresponding row and column are highlighted in red. 
Additionally, 
more information about the cell, including the total number and the average length of the sequences, is displayed in a tooltip.
Similarly, hovering over a bar in the bar charts highlights the entire column or row, and detailed information about the contained sequences is offered.
Moreover, zooming and panning are offered to help an analyst move around the matrix.
Three modes of selection are supported in Matrix View: row, column, or cell, by clicking the item while holding a modifier key. Two sets of event sequences (determined by a cell or the intersection of a row and a column) can be selected to feed to the Comparison View for further analysis.


\subsubsection{Design alternative}

\begin{figure}[!htb]
 \centering
 \includegraphics[width=\linewidth]{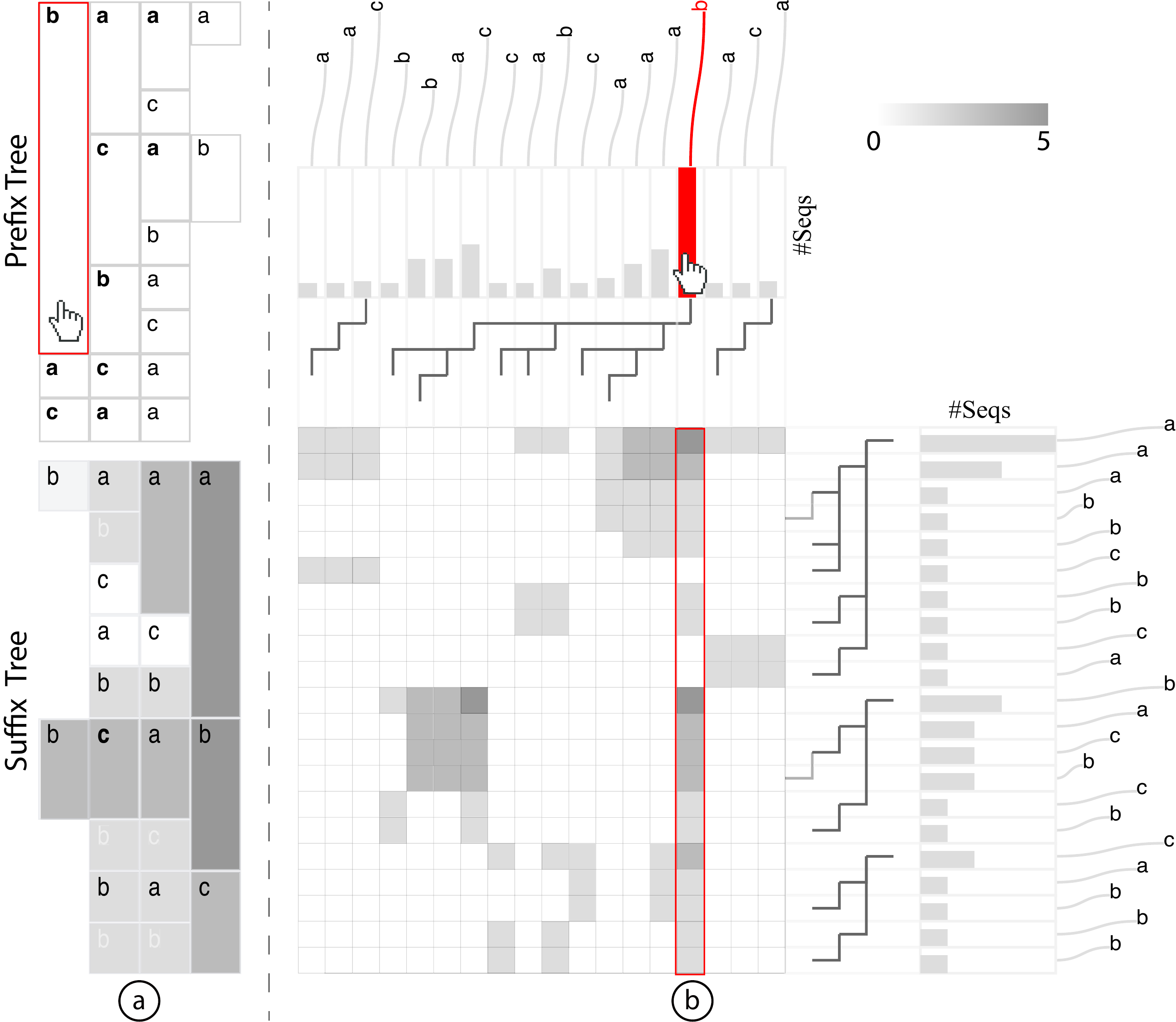}
 \caption{
 (a) An event sequence dataset is visualized with two icicle plots representing its prefix tree and suffix tree, respectively. When an analyst hovers over a node in one tree, nodes in the other tree are color coded by the number of associated sequences. For example, here ``b'' in the prefix tree is hovered over to investigate sequences starting with that event.
 (b) The same dataset is visualized with the matrix-based visualization. Even without a hovering over interaction, an analyst can clearly see that sequences starting with ``b'' are distributed widely in the suffix tree compared to those with other prefixes. 
 } \label{fig:designalt}
\end{figure}

During the design of \name, we explored a design based on two icicle plots~\cite{Kruskal1983}, which are placed side by side with each representing a prefix or a suffix tree of the data (Figure~\ref{fig:designalt}(a)).
The size of each node in icicle plots is mapped to the number of sequences it represents.
When an analyst hovers over a node in one icicle plot, the number of related sequences is mapped to the color of each node in the other icicle plot, as shown in Figure~\ref{fig:designalt}(a). 
Although this visualization is more space efficient, multiple hovering operations are needed to obtain a big picture about the distribution of event sequences with their prefixes and suffixes.
In addition, presenting multiple statistical information for each icicle node is not a trivial task. Hence, this design does not facilitate an analyst to choose interesting sets of sequences (\textbf{T2}).

As a result, we proposed the above matrix-based visualization that allows for more effective visual summarization and investigation of the entire dataset (Figure~\ref{fig:designalt}(b)). By employing various interaction techniques for expanding and collapsing, the matrix shows an overview of a dataset at multiple levels by balancing between space efficiency and information load. 

\subsection{Comparison View: distinguishing two sets}
After two meaningful sets of sequences are selected in the Matrix View, they are combined and fed into the sequential pattern mining algorithm~\cite{Fournier-Viger2014a}. 
The computed patterns, together with the raw sequences, are visualized in the Comparison View (Figure~\ref{fig:interface}(b)(d)) to support the visual comparison of the two sets at different granularities, including the pattern level (\textbf{T3}) and the sequence level (\textbf{T4}).

\subsubsection{Pattern level comparison}

As the aforementioned discussion, we employ the unit visualization technique for comparing two sets at the pattern level (\textbf{T3}). Each visual mark (\ie, a small rectangle) represents a sequence with its color indicating which set the sequence belongs to (Figure~\ref{fig:interface}(d)).
Multiple visual marks wrapped by a gray rectangle are displayed to indicate the pattern that these sequences contain, based on the layouts discussed in the design space.
In Section~\ref{sec:designspace}, we have identified two design alternatives, including ``Map2D\_MaxFill'' and ``FillX/Y\_MaxFill'', which are the most effective for the visual comparison at the pattern level in our case (Figure~\ref{fig:designspace}(i)(j)).
By default, the Comparison View uses the ``Map2D\_MaxFill'' configuration.
An analyst can switch among the two options by clicking the buttons on top of the Comparison View (Figure~\ref{fig:interface}(b)).
In addition, we provide options for choosing any combination of pattern and sequence layouts in Figure~\ref{fig:designspace} based on an analyst's experiences and goals.


\subsubsection{Sequence level comparison}

\begin{figure}[!tb]
 \centering
 \includegraphics[width=\linewidth]{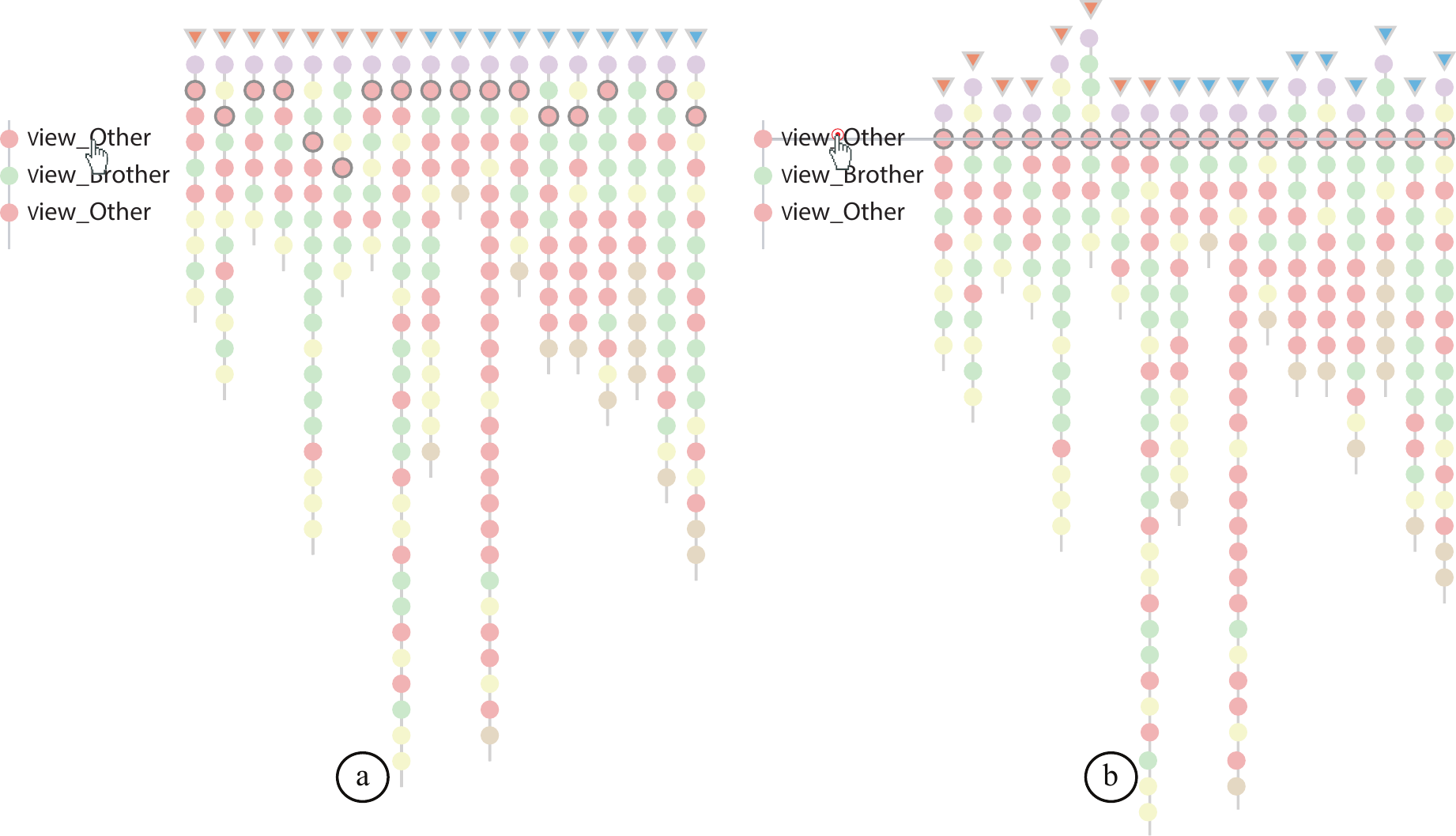}
 \caption{Sequence-level comparison. (a) Hovering over a key event highlights the corresponding event with gray border in each sequence. (b) Clicking a key event aligns all the sequences by that event.
 } \label{fig:sequence}
\end{figure}

After identifying patterns of interest, an analyst may want to examine how sequences in the support set manifest the pattern, and compare the two sets at the sequence level (\textbf{T4}). 
When an analyst clicks a pattern, all sequences in its support set are displayed on the right of the Comparison View (Figure~\ref{fig:interface}(f)).
Inspired by Liu et~al.'s work~\cite{Liu2017a}, we visualize the events of each sequence from top to bottom, and align all sequences horizontally.
Each event is represented as a circle with the color mapped to event type.
At the top of each sequence, a small triangle filled with blue or orange color is presented to indicate which set it belongs to. 

On the left of all sequences, key events in the pattern are displayed with text labels and aligned vertically in order. 
Several interactions are implemented to assist with the understanding of the relationships between the pattern and the corresponding sequences.  
First, when an analyst hovers over a key event, the first occurrence of such event is rendered with gray border in all sequences, as shown in Figure~\ref{fig:sequence}(a).
Second, when an analyst clicks a key event, sequences are aligned by this event (Figure~\ref{fig:sequence}(b)). Animated transitions are also applied to indicate the change of sequences in the alignment.


\section{Case studies}
\label{sec:casestudy}

In this section, we present three case studies from different domains to illustrate how analysts use \name{} to explore real-world datasets and obtain insights. 
The first dataset records important action events in football matches~\cite{football}, the second one tracks the interactions when students perform digital electronics exercises~\cite{Vahdat2015}, and the third one contains clickstreams of customers on an E-commerce website~\cite{ecommerce}. 


\subsection{Case I: analyzing football matches}

\begin{figure}[tb]
 \centering
 \includegraphics[width=\linewidth]{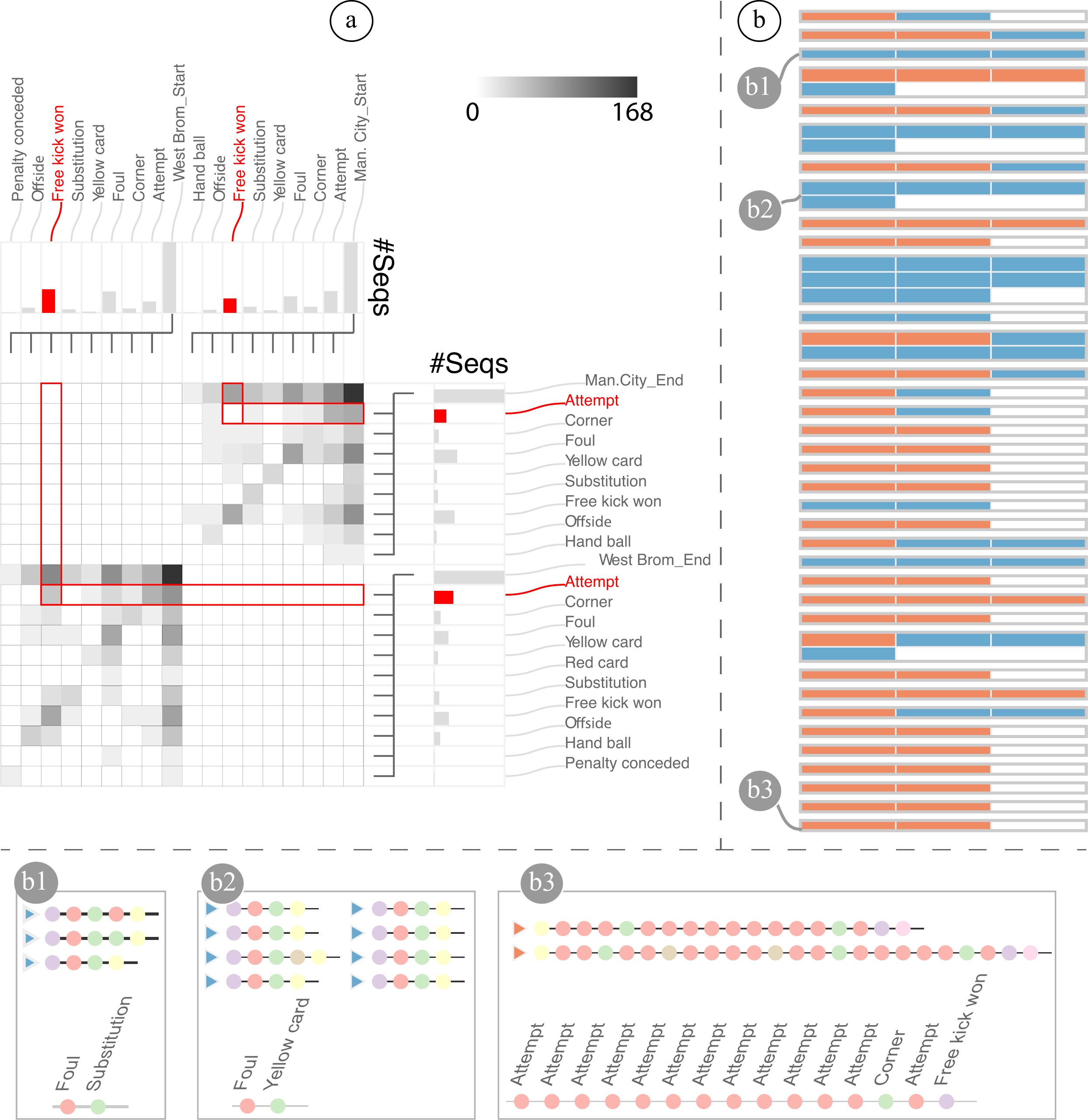}
 \caption{Tom is utilizing \name{} to analyze action event sequences of football matches between Man City and West Brom. (a) From the matrix, he observes that the cell (\textit{Free Kick Won}, \textit{Attempt}) from West Brom is gray, whereas the cell with the same events from Man City is white.
 (b) After selecting the two darkest cells (in orange and blue borders) in the matrix, Tom chooses the ``FillY\_MaxFill'' layout to compare patterns in these two cells in the Comparison View. 
 } \label{fig:football}
\end{figure}

The football match dataset records events in 7 matches between Manchester City (Man City) and West Bromwich (West Brom), where 626 events are categorized into 15 event types, including \textit{Announcement}, \textit{Attempt}, \textit{Corner}, \textit{Foul}, \textit{Yellow Card}, \textit{Second Yellow Card}, \textit{Red Card}, \textit{Substitution}, \textit{Free Kick Won}, \textit{Offside}, \textit{Hand Ball}, and \textit{Penalty Conceded}, \textit{West Brom Start}, \textit{West Brom End}, \textit{Man City Start}, \textit{Man City End}.
After sorting the events based on the timestamp, a sequence is defined as a list of events that is launched by one team before it is interrupted by the other. 
For example, during a ball control of West Brom, they may have a sequence like: \textit{West Brom Start}, \textit{Free kick won}, \textit{Foul}，and \textit{West Brom End}.
The whole dataset contains 355 sequences and the average length is $3.85$.

Assume Tom is a fan of West Brom and wants to know the behavioral difference between his favorite team and its opponent Man City that has a higher rank. 
To begin with, Tom wants to have a general idea about the event sequences of two teams by analyzing starting and ending events. 
Upon loading the dataset, Tom sees two dark cells in a $2\times 2$ matrix where columns are \textit{Man City Start} and \textit{West Brom Start}, and rows are titled \textit{Man City End} and \textit{West Brom End}.  
The two dark matrix cells, \ie, (\textit{Man City Start}, \textit{Man City End}) and (\textit{West Brom Start}, \textit{West Brom End}) include all the sequences of the two teams. 
To learn what the two teams do at first and at last during their turns, 
Tom clicks the two cells to expand both columns and rows to the second level (Figure~\ref{fig:football}(a)). 
He observes two clusters in the matrix representing Man City and West Brom, respectively.
With the thought that \textit{Free Kick Won} is a kind of starting point with various possible endings, 
he focuses on the columns titled with \textit{Free Kick Won} at the second level.
He observes that the cell (\textit{Free Kick Won}-\textit{Attempt}) of West Brom is darker in color.
However, the same cell of Man City is empty as shown in Figure~\ref{fig:football}(a).
This indicates that some attempts of West Brom start with getting free kicks, but no attempt is initialized by free kick for Man City. 
Thus, Tom guesses that West Brom tends to use the free kick as the beginning of a set play that leads towards an attempt opportunity.


Tom wonders if there is a key difference between the two teams during their ball control.
He selects the two darkest cells, \ie, 
(\textit{Man City Start}, \textit{Man City End}) and (\textit{West Brom Start}, \textit{West Brom End}), which are rendered with orange and blue borders in the matrix, respectively. 
Then sequential patterns are calculated and displayed in the Comparison View.
He explores various layouts of patterns and sequences by clicking the buttons on top of the Comparison View. 
There are some interesting insights shown in the ``FillY\_MaxFill'' layout when sorted by pattern length (Figure~\ref{fig:football}(b)).
He finds that the blue rectangles (West Brom) mainly appear on the top and the orange rectangles (Man City) scatter around the entire visualization. It indicates that some of the longest patterns are composed of the sequences of Man City only, and the sequences of West Brom are merely involved in shorter patterns.
Tom hypothesizes that Man City has better abilities to control the ball and clearer strategies in offense. 
Then, he clicks some of the longest patterns (appeared at the bottom of the Comparison View) and browses the key events on the right of the Comparison View (Figure~\ref{fig:football}(b3)). 
He observes that these patterns consist of continuous \textit{Attempt} with some \textit{Corner} and \textit{Free Kick Won}, indicating that Man City launches dense offense effectively when the ball is under their control. 
After clicking the patterns within the blue rectangles (Figure~\ref{fig:football}(b1, b2)), Tom identifies some key events including \textit{Foul}, \textit{Yellow Card}, and \textit{Substitution}.
Tom infers that West Brom faces great challenges in both defense and offense.


\subsection{Case II: understanding students' learning behaviors}

\begin{figure}[tb]
 \centering
 \includegraphics[width=\linewidth]{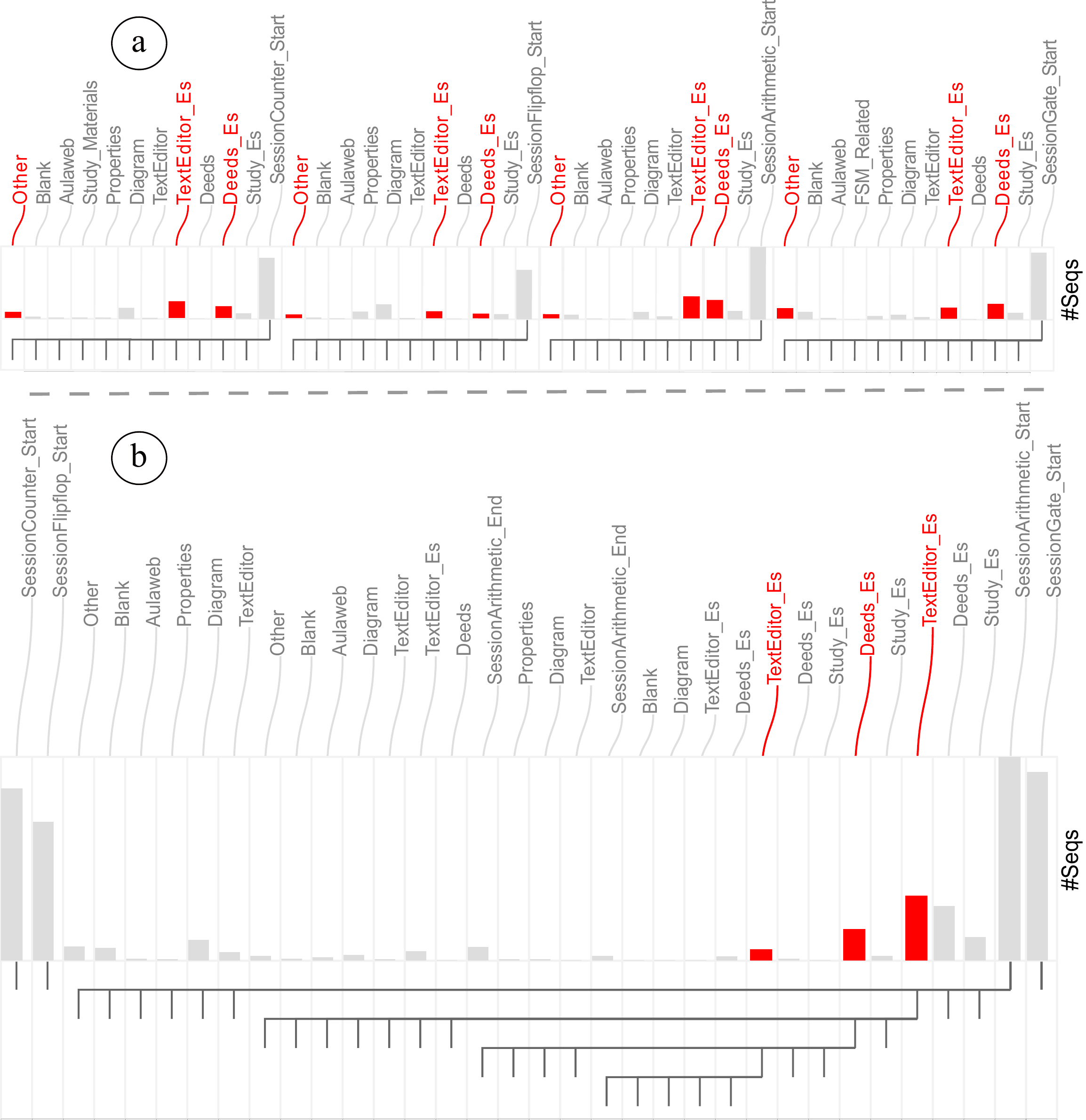}
 \caption{Linda is using \name{} to investigate student interactions using a course software called Deeds in four exercise sessions.
 (a) With the starting events of the sessions unfolded, the bar chart above the matrix shows that two events, \textit{TextEditor\_Es} and \textit{Deeds\_Es}, are dominant in the sequences.
 (b) The columns with the most numbers of sequences are iteratively expanded from the top level under session Arithmetic. 
 } \label{fig:exercises}
\end{figure}

Digital Electronics Education and Design Suite (Deeds) is a software for e-learning in digital electronics, which is used for digital circuit simulation, finite state machine simulation, and so on. 
When students do exercises of digital electronics with Deeds, the procedure usually involves sequences of events. For example, when a student starts to read the content of an exercise, the \textit{Study\_Es} event is recorded. Then, she may use Deeds to do exercises (\textit{Deeds\_Es}).
The student sometimes needs to draw diagrams (\textit{Diagram}) and adjust properties of the simulation (\textit{Properties}). 
Finally, she may use a text editor to write a report (\textit{TextEditor\_Es}), and do irrelevant activities to the course, such as open a web browser (\textit{Other}).
Overall, the dataset contains 8005 events with 23 event types. The events are recorded with a timestamp, a student ID, and a session ID. 

Suppose that students in the course Digital Electronics are required to finish four sessions of exercises with increasing difficulty, which focus on gates, arithmetic circuits, flip-flops, and counters, respectively.
Linda is a teaching assistant of the course and some students complain to her that switching between several software makes them easily distracted. 
So she wants to know if it is a common problem faced by all students and how students behave differently in different exercise sessions. 
She defines a sequence as a series of consecutive events performed by one student doing one exercise.
To facilitate the analysis based on sessions, each sequence starts and ends with events representing the session identifiers, such as \textit{SessionArithmetic\_Start},\textit{SessionArithmetic\_End}. 
Finally, she obtains 977 sequences with average length $5.04$.





First of all, Linda wants to explore the whole dataset to understand what students do during the exercise sessions. 
After loading the dataset, she filters the sequences with length smaller than three to remove outlier sequences. Then, she observes a $4\times 4$ matrix with the diagonal cells indicating the event sequences from the four different sessions.
Wondering what the most common starting events are in each session, she unfolds all the columns in the matrix (Figure~\ref{fig:exercises}(a)).
Linda observes that no matter what the session is, \textit{Deeds\_ES} and \textit{TextEditor\_Es} have relatively higher numbers of sequences based on the column bar chart, indicating that most students start their exercises by using Deeds and editing text. 
Further, Linda notices that the \textit{Other} bar under \textit{SessionGate\_Start} and \textit{SessionCounter\_Start} are higher than the other two sessions. 
With the preliminary knowledge that session Gate is the easiest and session Counter is the hardest, Linda hypothesizes that students may finish exercises in session Gate quickly and then do something irrelevant to the course like surfing the Internet (\ie, the event \textit{Other}). 
For exercises in session \textit{Counter}, which requires students to analyze counters consisting of many kinds of flip-flops, students may need to browse web-pages to search basic information online about these flip-flops.

Now Linda wants to confirm if students switch among different software. She focuses on the session Arithmetic because this session contains the most sequences as shown in the bar chart (Figure~\ref{fig:exercises}(a)). 
To drill down along the sequences and study the most common behaviors, she keeps expanding the columns with the most numbers of sequences progressively at different tree levels, under the session Arithmetic.
Linda observes that some sequences end with alternating between \textit{Deeds\_ES} and \textit{TextEditor\_ES} (Figure~\ref{fig:exercises}(b)), suggesting that students need to frequently switch between Deeds and a text editor. 
Linda further investigates other sessions and also finds such alternating behaviors. 
She concludes that it is common for students to switch among several software when doing exercises. 
Therefore, she would like to suggest the Deeds development team integrating a light text reader and text editor into the current system.


Next, Linda wonders if students perform differently in different sessions. 
Both bar charts next to the matrix columns and rows indicate that session Arithmetic and Flipflop 
contain the fewest and the most sequences among the four sessions, respectively. 
Thus, she selects these two cells for comparison, including (\textit{SessionArithmetic\_Start}, \textit{SessionArithmetic\_End})  and (\textit{SessionFlipflop\_Start}, \textit{SessionFlipflop\_End}), which represent all sequences from the two sessions (Figure~\ref{fig:interface}(c)). 

Then, Linda shifts her focus to the Comparison View with the ``Map2D\_MaxFill'' layout by default (Figure~\ref{fig:interface}(d)). 
She clicks the patterns containing only orange sequences which belong to the Arithmetic session, and discovers that they mainly consist of \textit{Deeds\_ES}, \textit{TextEditor\_Es}, and \textit{Study\_ES} (Figure~\ref{fig:interface}(e)). 
Since the Arithmetic session focuses on arithmetic circuit design, it  requires students to draw the schematics of circuits in the Deeds. 
Moreover, she observes that some patterns consisting of only blue sequences (from the Flipflop session) are clustered at the bottom left of the view (Figure~\ref{fig:interface}(d)). 
By clicking some of them, Linda discovers that these patterns contain \textit{Diagram} and \textit{Property} as key events (Figure~\ref{fig:interface}(f)). 
That is mainly because exercises on the Flipflop session requires students to frequently change input properties during simulations.

\subsection{Case III: investigating website clickstreams}

\begin{figure}[tb]
 \centering
 \includegraphics[width=\linewidth]{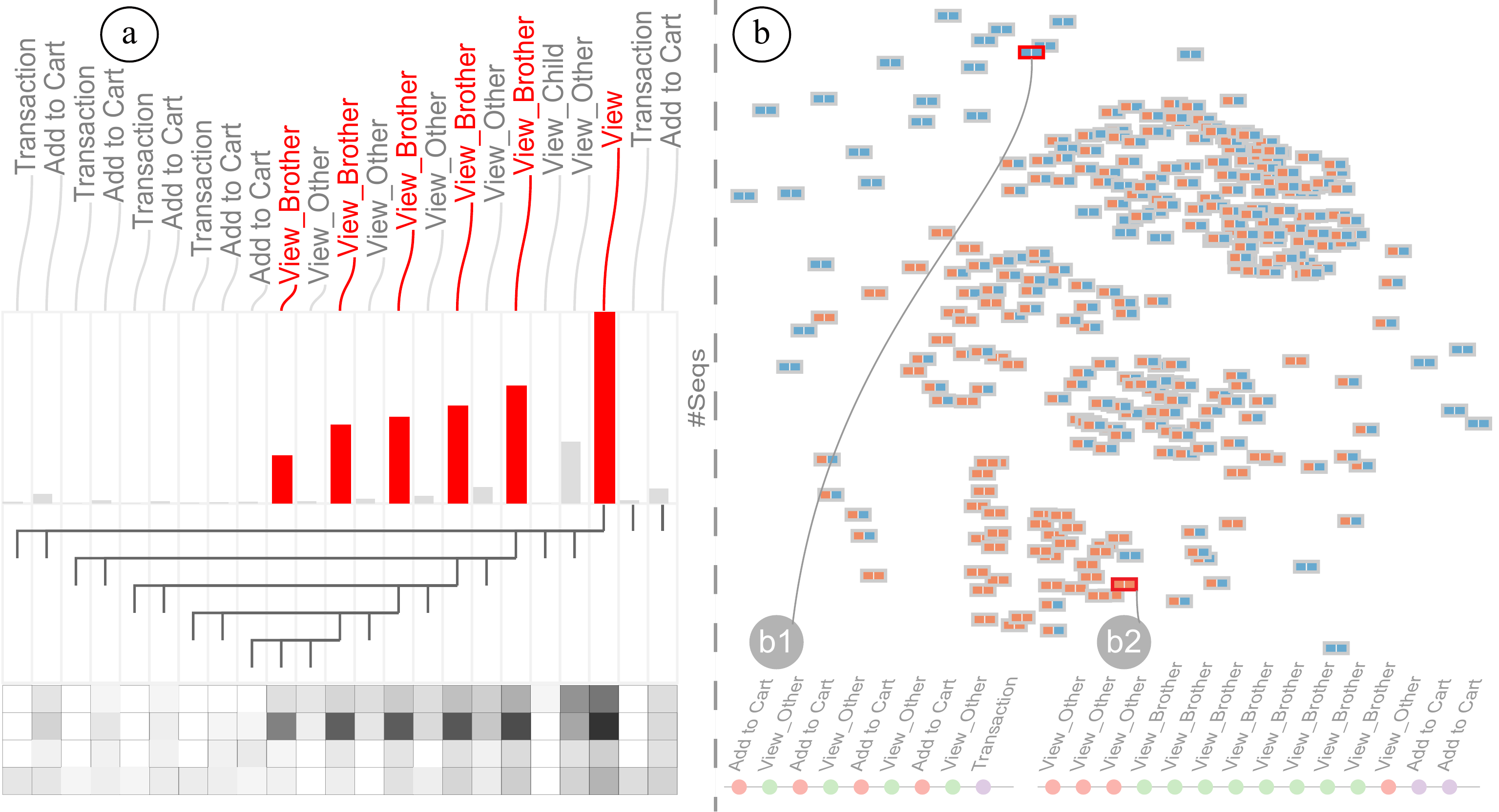}
 \caption{Sam is using \name{} to analyze clickstreams of customers on an E-commerce website.
(a) The columns titled with \textit{View\_Brother} are expanded iteratively till the fifth level.
(b) Two clusters of patterns, one mostly with blue sequences and one with orange ones, are observed in the Comparison View. 
 } \label{fig:ecommerce}
\end{figure}

This dataset collects customers’ clicks when they shop via a real-world E-commerce website. 
Customers can \textit{view} items, then \textit{add to cart}, and finally \textit{transaction}.
According to the category tree~\cite{ecommerce}, customers might \textit{view} items from the same, parent, or child categories, or the two items can be irrelevant from each other.
Hence, an event is defined by the relationship between the category of current viewing item and the previous one, such as \textit{View\_Parent}, \textit{View\_Child}, \textit{View\_Brother}, and \textit{View\_Other}. 
Overall, the dataset contains 4637 events with 7 types.
Thus, a sequence is a list of consecutive events made by one customer, which results in 820 sequences with the average length $7.65$.
Assume that Sam is a web analyst of this E-commerce platform and he would like to get some ideas about how customers shop on the this website.


Initially, the Matrix View shows a $4\times 3$ matrix with most cells colored black.
He first expands the column titled with \textit{View}, and notices that \textit{View\_Other} and \textit{View\_Brother} are dominant at the second level based on the column bar charts, and \textit{View\_Brother} is higher than \textit{View\_Other} (Figure~\ref{fig:ecommerce}(a)).
It indicates that most people start their exploration among items in the same category.
Sam keeps expanding \textit{View\_Brother} iteratively, and observes that the \textit{View\_Brother} bars are always the highest (Figure~\ref{fig:ecommerce}(a)), reflecting that the main event following \textit{View\_Brother} is still \textit{View\_Brother}. 
Next, he unfolds the \textit{View\_Other} column from the second level in the same way and finds that the most significant events following \textit{View\_Other} are \textit{View\_Other}. 
These observations show two distinct browsing behavior, which indicates two kinds of customers. 
One group tends to compare many similar items and may hope to find the best one; while the other one would like to view items from irrelevant categories. 


In addition, Sam observes an interesting phenomenon that some customers stop their visits by \textit{Add to Cart}, which means they do not checkout in the end. 
Thus, Sam would like to discover the difference between these customers. 
From the Matrix View, he selects two darker cells, (\textit{View}, \textit{Add to Cart}) and (\textit{View}, \textit{Transaction}).
and shifts his focus to the Comparison View, where all patterns are positioned by the ``Map2D\_MaxFill'' layout (Figure~\ref{fig:ecommerce}(b)). 
On the top left corner, Sam finds a cluster of patterns containing only the blue sequences from the cell (\textit{View}, \textit{Transaction}).
He clicks some of the patterns and finds that \textit{View\_Other} is the dominant event (Figure~\ref{fig:ecommerce}(b1)).
Meanwhile, a cluster of patterns on the bottom with many orange sequences from the cell (\textit{View}, \textit{Add to Cart}), attracts his eyes.
He explores these patterns and notices that \textit{View\_Brother} is the major event (Figure~\ref{fig:ecommerce}(b2)).
Sam infers that customers who explore similar items are those cannot make up their mind in buying things; on the contrary, those who have clear targets, usually switching between different categories, are likely to checkout finally.



\section{Discussion}
\label{sec:discussion}

Although the three case studies from various domains have demonstrated the effectiveness of \name{} in visually summarizing event sequence datasets and comparing sets of sequences at different granularities, the current prototype still has limitations.

First, although the Matrix View is scalable to the number of sequences, it may not scale well when the cardinality of the event sequence data is large. This may result in a large matrix visualization as the types of events are too many. 
Also, the node expanding and collapsing operations on the matrix columns and rows may overwhelm analysts when the depth of trees is larger.
However, categorizing events based on their attributes might solve this problem.

Another limitation of the matrix is regarding its capability in selecting sequence sets with more complicated criteria. The current \name{} design focuses on the identification of sequences according to their prefixes and suffixes. In future research, we plan to explore visualizations that allow for identifying sequences based on event attributes, timestamps, time intervals, etc. 
Further, supporting event querying specifications beyond just prefix and suffix, such as using regular expression like in (s$\vert$qu)eries~\cite{Zgraggen2015}, would greatly empower the flexibility of the matrix visualization. That is, each matrix row or column could be a specification and each cell represents all the sequences satisfying both criteria from the row and the column. 

Third, unit visualization may have three main disadvantages, including computational scalability, display scalability, and perceptual scalability~\cite{Park2017}. The Comparison View shares all these scalability issues, but they can be addressed by adjusting the computation process from two aspects. 
On one hand, an analyst can choose to perform random sampling for the entire dataset in the Matrix View to limit the volume of input data. 
However, random sampling may not be the ideal option because it ignores the characteristics of the datasets, \eg, the distribution of sequences.
On the other hand, an analyst can adjust the ``support'' value for the maximal pattern mining algorithm~\cite{Fournier-Viger2014a} in the Comparison View, to obtain fewer patterns in general.
However, this prevents an analyst from ``seeing'' all the patterns that may result in missing opportunities.

There also exists limitations of our study. Although we have applied \name{} to three real-world datasets to demonstrate its generalizability to various domain applications, deployment studies are required to further verify our conclusions. Further, controlled user studies need to be conducted to validate our choices and to better understand the pros and cons of each design alternative in Figure~\ref{fig:designspace}.

There are a number of promising interesting directions to extend our visualization techniques. 
First, the Matrix View can be applied to other tasks, such as the origin-destination analysis. The multi-level matrix can demonstrate the traffic volume at different scales. 
\siwei{We notice that MapTrix~\cite{Yang2017} employs a matrix to show the flow volume between origins and destinations. However, it does not support exploration at multiple levels.}
Second, some of our discussion of design space can be generalized to guide other work in applying the unit visualization technique to visual comparison, 
such as the advantages and disadvantages of combination of two layouts in comparison tasks.
Since we only discuss one configuration in unit visualization, \ie, pattern layouts are uniform while sequence layouts are non-uniform, we plan to address other configurations in the future. 


\section{Conclusion and future work}
We have introduced an interactive visualization, called \name{}, for summarizing an event sequence dataset based on prefixes and suffixes, and further helping analysts identify promising sets of event sequences to compare at both the pattern and sequence levels. To design \name{}, we have performed tasks analysis based on Brehmer and Munzner's typology~\cite{Brehmer2013}. 
To support the comparison task, we have explored the design space of employing the unit visualization technique in our scenario.
Moreover, we have described three case studies to illustrate the effectiveness and usefulness of \name{} with real-world datasets in three different application domains.

In the future, we plan to enhance the matrix-based visualization by encoding attributes, such as event timestamps and time intervals, on the columns and rows, as well as by supporting more complicated event specifications in addition to just prefix and suffix.
Thus, analysts are able to identify and select interesting sets of sequences based on both event attributes and cardinality in a richer manner.
Next, to verify the hypotheses in our discussion for the design space (Figure~\ref{fig:designspace}), we plan to conduct empirical studies to evaluate user performance with different design alternatives.
Finally, we are interested to experiment \name{} with other comparison tasks in different application domains.

\acknowledgments{
We thank all the domain experts involved in the studies. We also thank Miss Du Rao for helping us record the video. 
} 

\bibliographystyle{abbrv-doi}

\bibliography{sample}
\end{document}